\documentclass[twocolumn,prstab]{revtex4-1}    
\usepackage{graphicx}

\begin{document}
\title{Performance of solenoids vs. quadrupoles in focusing and energy selection of laser accelerated protons.}

\author{Ingo Hofmann}
 \altaffiliation[Also at ]{Helmholtz-Institut Jena,
Helmholtzweg 4, 07743 Jena, Germany}
\affiliation{GSI Helmholtzzentrum f\"{u}r Schwerionenforschung
GmbH Planckstrasse 1 64291 Darmstadt, Germany}
\email{i.hofmann@gsi.de}


\begin{abstract}

 Using laser accelerated protons or ions for various applications - for example in particle therapy
 or short-pulse radiographic diagnostics -
 requires an effective method of focusing and energy selection. We derive an analytical
 scaling  for the performance of a solenoid compared
 with a doublet/triplet as function of the energy, which is confirmed by TRACEWIN simulations.
  The scaling shows that above a few MeV a solenoid needs to be pulsed or super-conducting, whereas
  the quadrupoles can remain conventional. The transmission of the triplet
  is found only 25\% lower than that of the equivalent solenoid. Both systems are equally suitable for energy selection
  based on their chromatic effect as is shown using an initial
  distribution following the RPA simulation model by Yan et al.\cite{yan2009}.

\end{abstract}

\maketitle \vspace{1cm}

\setcounter{page}{1}
\section{INTRODUCTION} \label{sec:1} Laser acceleration of protons or ions requires
ultra high laser intensities focused on thin target foils. This
has been demonstrated in numerous experiments (see, for instance,
Refs.~\cite{snavely,mckenna,hegelich,gaillard}). Applications of
this novel acceleration method have been suggested, for example in
terms of a new proton source for radiation
therapy~\cite{bulanov2002,fourkal2002,malka2004} as alternative to
more conventional accelerator technologies, like protons or ions
from cyclotrons or synchrotrons. Other potential applications
might be proton radiography, neutron imaging or isotope
production. Energies up to 170 MeV for deuterons have recently
been observed  at the TRIDENT laser and explained as ``break-out
afterburner'' (BOA) mechanism~\cite{roth2013}, with possible
applications as very short-pulse neutron source.

All of these applications have to cope with the characteristics of
laser accelerated protons or ions: a large energy spread as well
as angular spread, sub-ns time scales, significant shot-to-shot
fluctuations and - for practical applications - relatively low
repetition rates compared with conventional accelerators. This
requires specific methods to suitably manipulate laser accelerated
protons in space and time and match them to the need of an
experiment or application.

 In a preceding study we have shown that a single solenoid
magnet can be used very effectively to combine  angular focusing
(collection) with energy selection due to the lens chromatic
effect~\cite{hofmann2012a}. This is extended here to a comparative
evaluation of quadrupole focusing (doublet or triplet) and
solenoid focusing. As in the solenoid case, we use the dependence
of focal length on energy and employ a radially confining aperture
to select a suitable energy window. The main difference of
quadrupoles versus a solenoid is their first order focusing
property (solenoids focus in second order) and the asymmetry in
focusing (astigmatism) as well as different chromatic effects
between the horizontal and vertical planes.  On the other hand
alternating gradient focusing with quadrupoles is known to be more
efficient at increasing energy and helps to avoid pulsed magnet or
superconducting  technology.

It should be mentioned here that the alternative to energy
selection by the chromatic effect is the more conventional energy
selection using the dispersive properties of dipole magnets. All
therapy oriented studies on laser acceleration carried out in the
course of the last years have used such a system with a
collimating aperture followed by a dispersive dipole
magnet~\cite{Ma,PMRC,dresden1,dresden2}. Dispersive energy
selection is certainly an option, but schemes without transverse
focusing  may result in significant efficiency loss. Chromatic
energy selection, instead, combines the advantages of focusing
with energy selection in a single device. However, for effective
selection it is required that the beam size by chromaticity
dominates over the size by the intrinsic emittance, which is   not
the case in many situations - but certainly for laser accelerated
particles.

An interesting alternative is the micro-lens, which focusses by
means of an electron plasma initialized by a second short-pulse
laser beam and collapsing inside a cylinder~\cite{toncian}. Energy
selection is enabled by using time-of-flight and triggering the
second laser beam accordingly. For therapy applications, however,
where energy selection failure cannot be tolerated, the fully
controllable powering of conventional magnets for energy selection
seems essential.

In Section~\ref{sec:2} we compare solenoidal with doublet/triplet
focusing. Their chromatic focusing properties are studied in
Section~\ref{sec:3}. In Section~\ref{sec:4} results are applied to
energy selection by an aperture for both systems using an input
distribution from the simulation of a radiation pressure
acceleration model. In Section~\ref{sec:5} we draw conclusions.

\section{Comparison of solenoidal and quadrupolar
focusing}\label{sec:2} Solenoids are frequently applied in
injectors for focusing of particles with relatively low energy and
large divergence. Likewise, focusing of laser accelerated protons
with energies of the order of 10 MeV  was demonstrated using a
pulsed solenoid~\cite{nuernberg,Burris}. Other laser proton
experiments in a comparable energy range have successfully
employed small aperture, high-gradient permanent magnet
quadrupoles~\cite{schollmeier}. Preference of quadrupoles over
solenoids depends on the individual application, but certainly
energy and the question of room-temperature, non-pulsed
quadrupoles versus pulsed or superconducting solenoids matter.

A useful guidance to decide on the basis of required field
strengths can be obtained from a scaling expression using the thin
lens approximations for the focal length $f_s$ of a solenoid and
$F_d$ of a quadrupole doublet as suggested in
Ref.~\cite{reiserbook}.  With $B$ the field strength (for the
quadrupoles defined at the poles), and assuming that both focal
lengths are defined from the respective centers, we have for the
solenoid of length $L$
\begin{equation}\label{sol}
1/f_s \approx(\frac{q}{2mc\beta\gamma})^2B^2L;
\end{equation}
 likewise for the doublet
\begin{equation}\label{doub}
1/F_d\approx (\frac{qB}{mc\beta\gamma a})^2 l^2s,
\end{equation}
where $l$ is the individual quadrupole length, $s$ the separation
of quadrupoles (from center to center) and $a$ the maximum beam
radius (pole radius). Note that the focal strength of a doublet
increases with the separation of its components - of course on the
expense of decreasing acceptance. Comparing  a solenoid with a
doublet of the same overall length $L$ and equal field $B$,  we
readily obtain from Eqs.~\ref{sol} and  \ref{doub} the ratio $T_d$
of focusing strengths (here defined as inverse focal lengths) in
terms of only geometrical quantities:
\begin{equation}\label{reiser}
T_d\equiv \frac{1/F_d}{1/f_s}=\frac{4sl^2}{a^2L},
\end{equation}
Eq.~\ref{reiser} indicates that the focusing strength of a doublet
is superior to that of a solenoid, if $a$ is sufficiently small
relative to the length. As an example, consider a doublet with a
gap between magnets equal to their length, in which case we have
$T_d=(2/3)^3(L/a)^2$ and the transition condition $T_d>1$   occurs
for $L/a>(3/2)^{3/2}$.

In order to obtain systems with equivalent focal lengths it may be
required to adjust $B_s$ according to Eq.~\ref{sol}. This results
in an effective field for the solenoid
\begin{equation}\label{equiv}
B_s^*=T_d^{1/2}B_d,
\end{equation}
 which may lead to the requirement of super-conducting or pulsed power technologies for the
 solenoid.

\begin{figure}[h]
\centering\resizebox{0.45\textwidth}{!}{\includegraphics{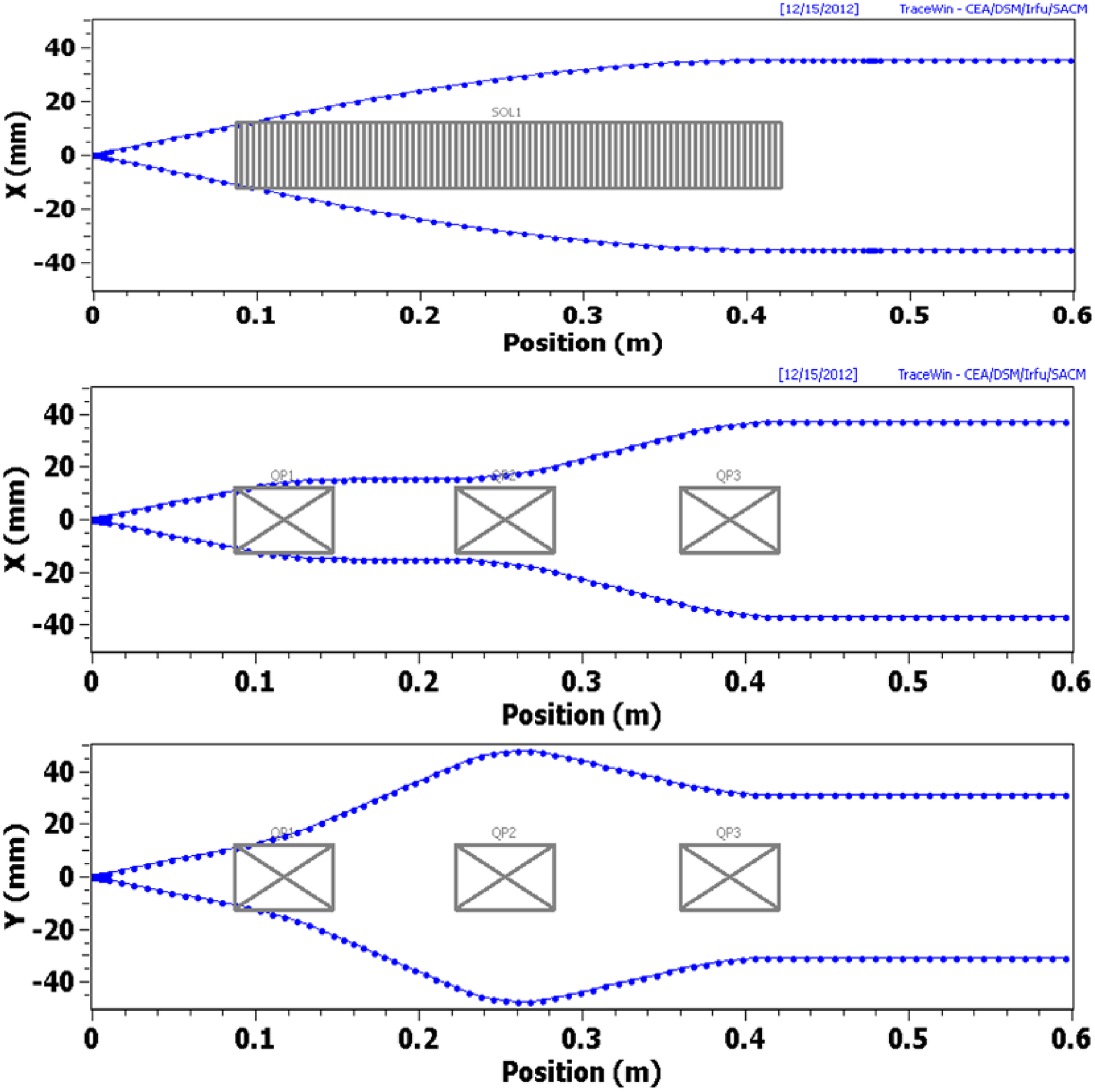}}
\caption{TRACEWIN envelopes for equivalent triplet (top) and
solenoid (bottom) solutions at 2 MeV.} \label{tripletsol}
\end{figure}

We can apply this to the collection of laser particles, if we
assume the focal spot is at the source and the beam is to be made
parallel by the lens.
 In the interest of smoother focusing we find it preferable to use a triplet rather than a doublet as reference case.
 In Fig.~\ref{tripletsol} we show - as example - a triplet with $L=0.334$ m, $l=0.06$ m and gradients of $30$ T/m, $-30$ T/m
 and $15$ T/m. The calculation of matched beam optics is obtained using the envelope option of the
 TRACEWIN code~\cite{tracewin}, which
  is also employed further below for particle tracking.

 2 MeV protons with source divergence of $\pm 125$ mrad are made parallel with a triplet
focal length
 (source to triplet center) $F_t=$0.254 m, maximum envelope $a=48$ mm, hence maximum pole-tip field $B_t=1.44$ T.
 Note that the distance
 source to lens is given by $F_t-L/2$, which is 87 mm in this
 case.
 The energy spread is assumed to be zero,
 which is to avoid the chromatic energy effect at this point. For comparison we also show a solenoid focusing with the
 same $L$; in order to achieve the same focal length we require a slightly increased magnetic field $B_s=1.53$ T.
Next we check if our TRACEWIN results still obey the scaling of
Eq.~\ref{reiser}, although we have replaced the doublet by a
triplet. We find  good agreement,  if we replace $4s$ by $2s$. The
need for this adjustment can be interpreted as necessary
compensation of the (approximate) doubling of the triplet length
compared with that of the doublet. This suggests a triplet
focusing enhancement factor
\begin{equation}\label{reiser1}
T_t\equiv \frac{1/F_t}{1/f_s}=\frac{2sl^2}{a^2L}.
\end{equation}

 Assuming that the
gap equals the quadrupole length, we have $s= 2l$, which results
in $T_t\approx 1.12$ in the above example. Applying
Eq.~\ref{equiv} we find that the predicted solenoid field for
equivalent focal length is 1.52 T, which agrees quite well with
the above TRACEWIN result - in spite of the thin lens
approximations employed in the derivation of $T_t$.

\begin{figure}[h]
\centering\resizebox{0.45\textwidth}{!}{\includegraphics{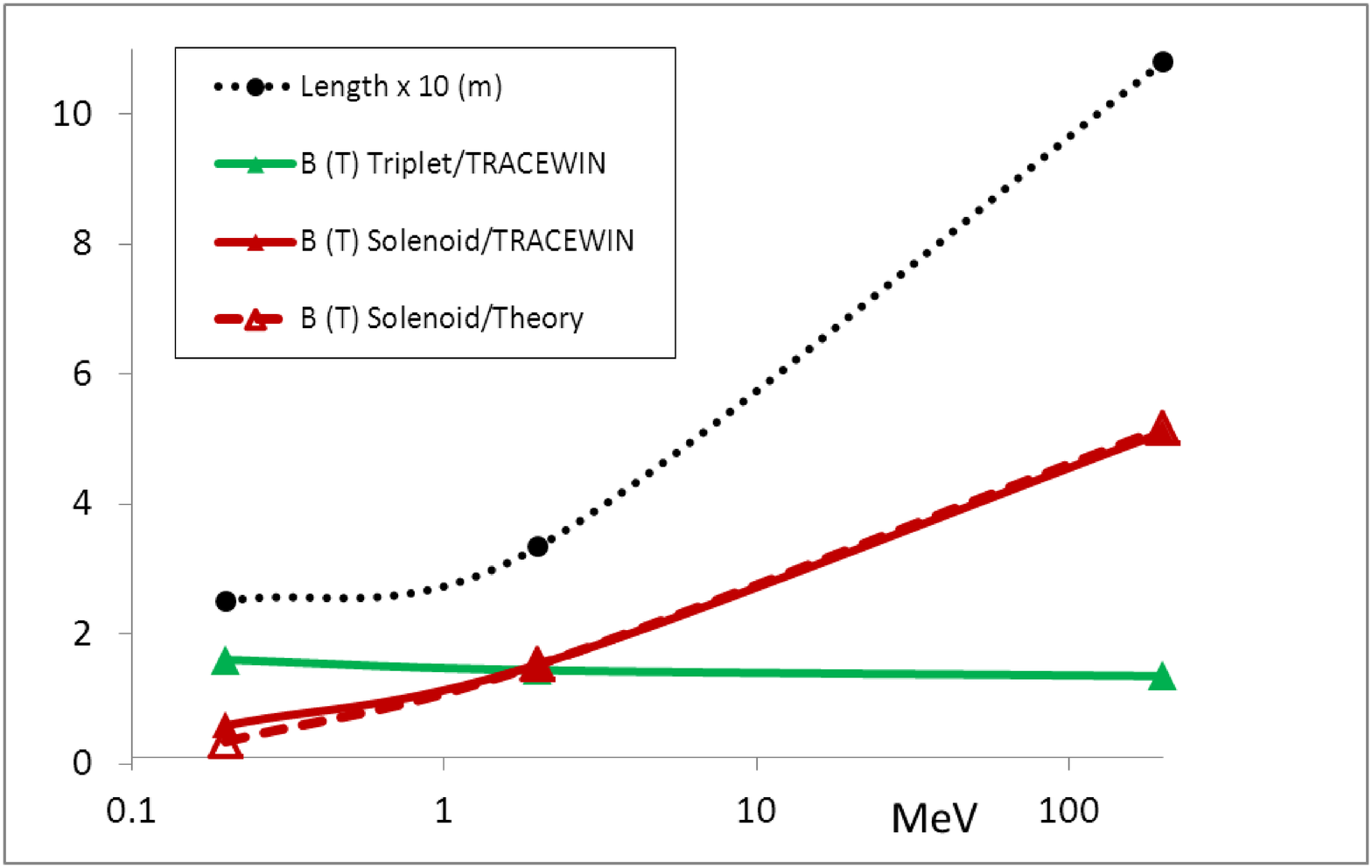}}
\caption{Comparison of  triplet and solenoid systems with equal
focal lengths as function of energy.} \label{scaling}
\end{figure}

In order to further examine the validity of Eq.~\ref{reiser1} for
different energies we extend the systems in Fig.~\ref{tripletsol}
to proton energies of 0.2, 20 and 200 MeV, again assuming at each
energy equal overall lengths $L$ for solenoid and triplet, equal
focal lengths for the two systems as well as equal gap and
quadrupole lengths in the triplet case. For the initial divergence
we assume - somewhat arbitrarily - a divergence scaling $x'\propto
(\beta \gamma)^{-1/2}$ to account for the expected trend of
decreasing divergence with energy. The value of $x'$ at 2 MeV is
kept as before. We first use TRACEWIN and search again for matched
solutions requiring a parallel output beam for vanishing energy
spread. In Table~\ref{table} we summarize all relevant parameters
including the resulting quadrupole pole tip fields B$_{t}$ and
solenoid strengths B$_{s}$. The focal length F is again defined
from source to center of the respective lens system and found to
increase with energy. The theoretically expected triplet
enhancement factor T$_t$ is calculated from Eq.~\ref{reiser1} ($s=
2l$) by inserting the respective geometrical dimensions. Using
Eq.~\ref{equiv}, which applies equally to the pole tip field of
the triplet as it does for the doublet, we can thus derive the
theoretically expected B${_s}^*$ and compare it with the actual
B$_{s}$ obtained from TRACEWIN matching.
\begin{table}[h]\label{table}
\begin{tabular}{|l|c|c|c|c|c|c|c|c|c|}
  E (MeV)&x'   &L    & F      & l   & a   & B$_{t}$    &B$_{s}$  &T$_t$  &  B$_{s}^*$\\ \hline
  0.2    & 400 &25.0 & 22.5   & 2   & 5.3 & 1.60       & 0.59    & 0.046 &  0.34  \\
  2      & 125 &33.4 & 25.4   & 6   & 4.8 & 1.44       & 1.53    & 1.12  &  1.52   \\
  20     & 71  &62.0 & 47.0   &10   & 4.8 & 1.50       & 2.62    & 2.8   &  2.51 \\
  200    & 39  &108.0& 79.0   &20   & 4.5 & 1.35       & 5.10    & 14.6  &  5.16 \\
\end{tabular}
\caption{Comparison simulation - theory for equivalent solenoid
and triplet focusing properties; lengths in cm and magnetic fields
in T ($x'$ in mrad).}
\end{table}
  As result we find  an overall good agreement between the TRACEWIN calculated B$_{s}$ and the theoretical
  B$_{s}^*$, which confirms the theoretically derived triplet focusing enhancement over a
solenoid. In Fig.~\ref{scaling} we
  summarize the main findings from this comparison.

In summary, this demonstrates that for sub-MeV or few MeV energies
solenoids are a convenient approach, whereas the quadrupole
doublet/triplet (or multiplet) has advantages  for higher energies
  as its pole tip field strengths remain within iron saturation.
Eq.~\ref{reiser1} also suggests that $T_t\propto 1/a^2$, hence
larger $a$ due to an increased initial divergence shifts the
advantage of the doublet/triplet over the solenoid to higher
energies - and vice versa.

\section{Chromatic properties of solenoid and triplet}\label{sec:3}

TRACEWIN is used here for particle tracking. Although primarily a
linear accelerator design and verification tool, it has a number
of features, which make the code suitable for our problem as well.
In particular, it
\begin{itemize}
    \item is a self-consistent 3D particle-in-cell code suitable for tracking of
    up to $10^7$ simulation particles,
    \item is capable of energy dependent focussing (chromatic
    aberrations),
    \item includes  field map options for magnetic elements (like solenoids) to
    model higher order (for example geometric) lens aberrations,
    \item provides standard 6D phase space initial distributions as well
    as user provided input distributions; as ``standard uniform" we choose for this study the option
    of a uniform distribution in
    the 4D transverse space space as well as uniform within the longitudinal phase plane ellipse;
    \item includes an envelope option for beam optics design.
    \end{itemize}
Space charge options with 2D/3D Poisson solvers exist, but space
charge is ignored in this study. In Ref.~\cite{hofmann2012a} it is
shown for solenoids that space charge is generally weak; in  the
near-source region, where extremely high proton densities are
prevailing, neutralization by the co-moving electrons helps.

 In the following we use as reference a solenoid and an equivalent
triplet, both 104 cm long and designed to bring 250 MeV protons to
a focus at 2.73 m with the following assumptions:
\begin{itemize}
  \item initial maximum divergence angle: $\pm$ 28 mrad
  \item energy spread: practically mono-energetic
    \item distance laser target - first magnet: 35 cm
     \item beam pipe radius: 3.5 cm
    \item aperture radius of solenoids and quadrupoles: 3 cm
    \item length of solenoid field map: 104 cm
    \item length of solenoid field region: 80 cm
    \item averaged solenoid field 6.27 T
    \item length of solenoid field map: 96 cm
    \item length of quadrupoles: 15 cm
    \item quadrupole pole tip fields: 1.5/1.5/1.0 T
 \end{itemize}

\begin{figure}[h]
\centering\resizebox{0.45\textwidth}{!}{\includegraphics{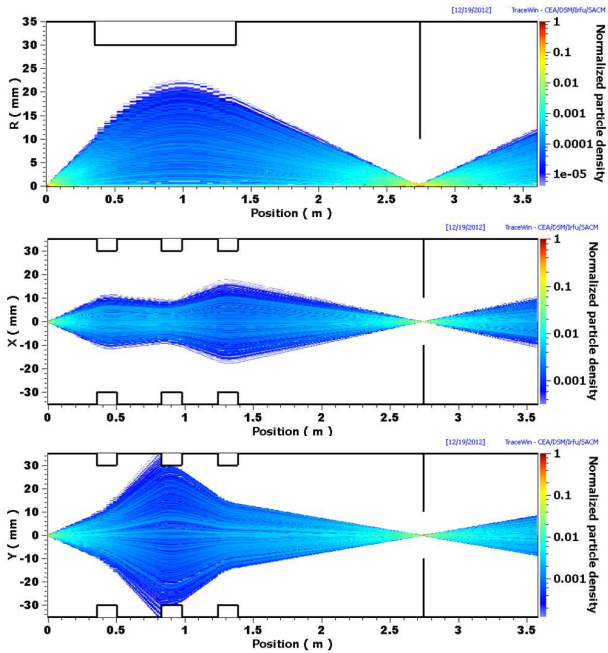}}
\caption{Density plots for TRACEWIN obtained reference cases for
``equivalent'' solenoid (top) and triplet (center:x and bottom:y)
focusing.} \label{sol-triplet}
\end{figure}

Fig.~\ref{sol-triplet} shows density plots from a multi-particle
simulation using a low number of simulation particles (only 3000),
which helps to visualize single particle rays. The simulation was
carried out with the ``standard uniform'' initial distribution of
TRACEWIN.  Maximum energy deviations in this example have been
chosen as $\pm 5 \times 10^{-4}$\ MeV centered at 250 MeV, hence
practically mono-energetic.  The common waist for $x$ and $y$ for
the triplet (stigmatic image) is relevant for optimum energy
selection as will be shown in the next section.
It is noticed that the de-focusing effect of the first quadrupole
in $y$ leads to a - in this example - small beam loss at the
aperture of the second quadrupole.

In Refs.~\cite{hofmann2012a}  the energy dependence of the focal
length of a solenoid lens was expressed in terms of a chromatic
coefficient, which we generalize here to cope with the different
focusing in $x$ and $y$ for a triplet:
\begin{equation}
\alpha_{x,y} \equiv \frac{\delta f_{x,y}/f_{x,y}}{\delta E/E}.
\label{chromatic-coefficient}
\end{equation}
Here $f_{x,y}$ is the focal length at the reference energy $E$ and
$\alpha_{x,y}$ is specific to the geometry of the focusing set-up.
For the examples of Fig.~\ref{sol-triplet} we find from TRACEWIN
simulation for the solenoid $\alpha_r\approx 1.9$ and for the
triplet $\alpha_{x}\approx 0.9$ as well as $\alpha_{y}\approx
3.8$. The much larger $\alpha_{y}$ is a result of the de-focusing
in $y$ at the first lens and the thus much larger overall envelope
excursions in $y$. The solenoid chromatic coefficient is - not
surprisingly - close to the geometrical mean of the two
coefficients for the equivalent triplet.

The  chromatic effect is strongly correlated with time as higher
energy particles travel ahead. The stronger focusing for lower
energies leads to an enhanced transverse phase space rotation of
protons at the bunch end as compared with the high energy
particles at the bunch head. The resulting slip in the transverse
phase planes causes an effective transverse emittance increase,
which can significantly exceed the initially small production
emittance. The effective emittance is obtained by averaging the
instantaneous emittances over the full bunch length, hence the
full energy spectrum~\cite{hofmann2012a}.

\section{Transmission and energy selection}\label{sec:4}
As suggested in  Refs.~\cite{hofmann2012a} the pronounced
chromatic focusing effect can be used for an effective energy
selection, if the beam is focused into a suitably defined
transverse aperture. Only particles with focal spot sufficiently
close to the aperture plane are transmitted effectively. As for
solenoids this works effectively only if the beam is
``chromaticity dominated'': at a selection aperture the beam size
by chromaticity dominates over the size generated by the intrinsic
emittance at any relevant value of the energy (i.e. position along
the bunch). This is always the case for laser generated ions with
their extremely small  emittance at any given energy, which is
owed to the very small source spot size. It should be mentioned
here that the intrinsic emittance should also include emittance
increase due to higher order aberrations of the lens, which can be
a problem for short solenoid lenses at low energy but is not
further considered in this study.

\subsection{RPA generated initial distribution}\label{RPA}
\begin{figure}[h]
\centering\resizebox{0.45\textwidth}{!}{\includegraphics{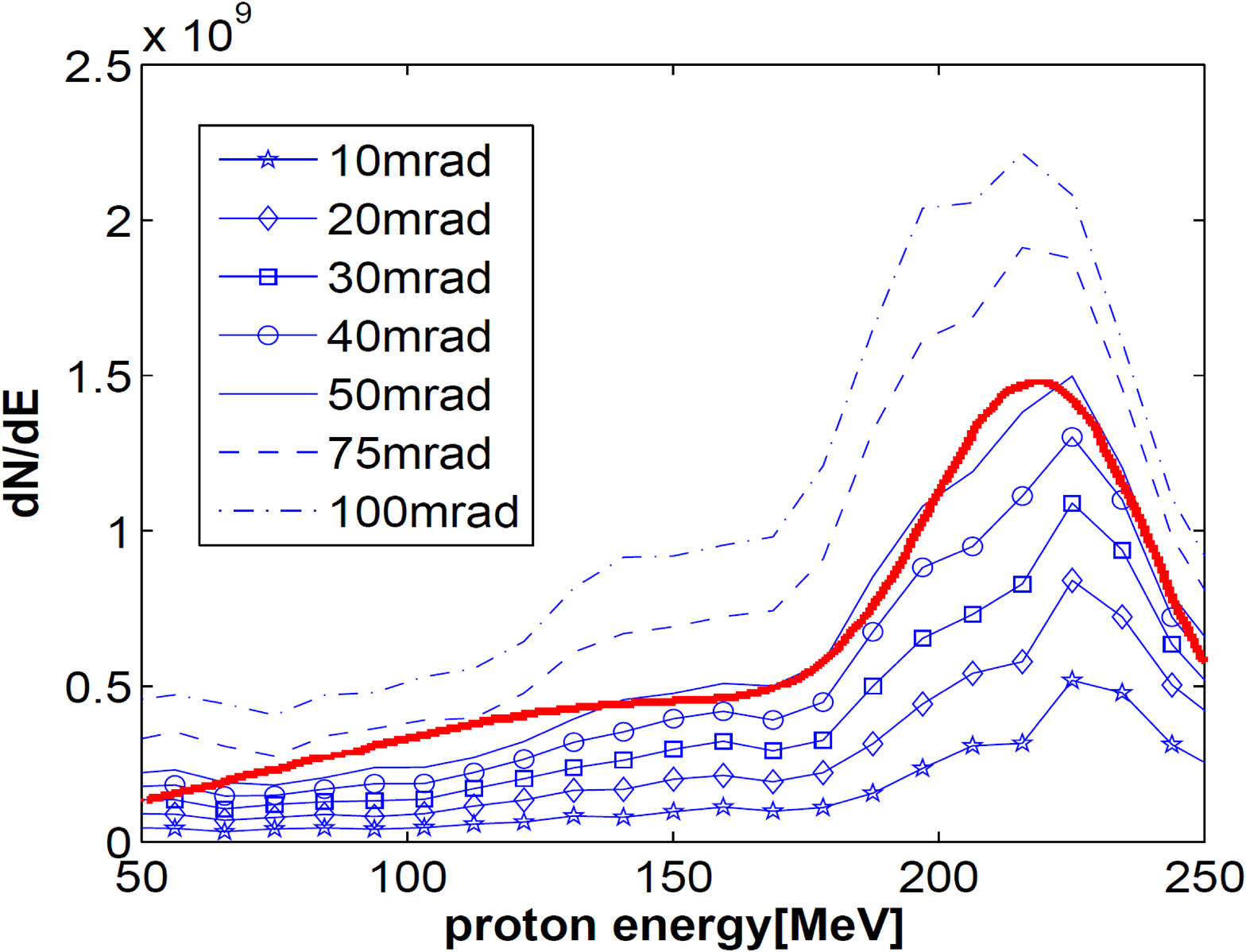}}
\caption{Spectral yield of protons as a function of  energy   and
for different capture cone angles $\Omega$, with bi-Gaussian fit
(continuous line).} \label{proton-spectra}
\end{figure}

For practical considerations it is advantageous to consider an
initial distribution in 6D phase space with a broadened energy
distribution according to some laser acceleration model. The
Radiation Pressure Acceleration (RPA)
mechanism~\cite{klimo,zhang,macchi,esirkepov04,henig2009,henig2009a}
has a high potential to reach proton energies of hundreds of MeV.
A specific theoretical version of it has been discussed in
Ref.~\cite{yan2009} and applied to proton therapy conditions in
Refs.~\cite{hofmann2011,hofmann2012a} to create a proton energy
spectrum extending up to 250 MeV. The output of this
RPA-simulation can be described as spectral yield
\begin{equation}\label{spectralyield}
    \frac{dN(E,\Omega)}{dE} [MeV^{-1}],
\end{equation}
which describes the number of particles in an energy interval $dE$
and within a cone angle $\pm\Omega$. The thus defined proton
spectrum (details see Ref.~\cite{hofmann2011})  is plotted in
Fig.~\ref{proton-spectra}. Its energy distribution is peaked above
200 MeV - determined by the laser intensity - with a relatively
broad foot towards lower energies. As input into our TRACEWIN
simulations we take a bi-Gaussian approximation to this energy
spectrum shown by the continuous curve in
Fig.~\ref{proton-spectra}.  The 6D initial distribution is taken
as a Gaussian random distribution in the variables $t, x, x', y,
y'$. For the rms widths in $x', y'$ we have chosen 35 mrad - in
contrast with the broader tails in divergence indicated in
Fig.~\ref{proton-spectra}, which are probably due to the 2D nature
of the RPA simulation. The initial spot radius and pulse duration
are in the $\mu$m rsp. ps scales; their actual values play no role
as long as space charge is considered as neutralized initially. We
also note that the detailed profile of the energy spectrum is only
exemplary - what matters primarily is its gradient near a selected
energy.


\subsection{Comparative transmission}\label{rpa}
\begin{figure}[h]
\centering\resizebox{0.45\textwidth}{!}{\includegraphics{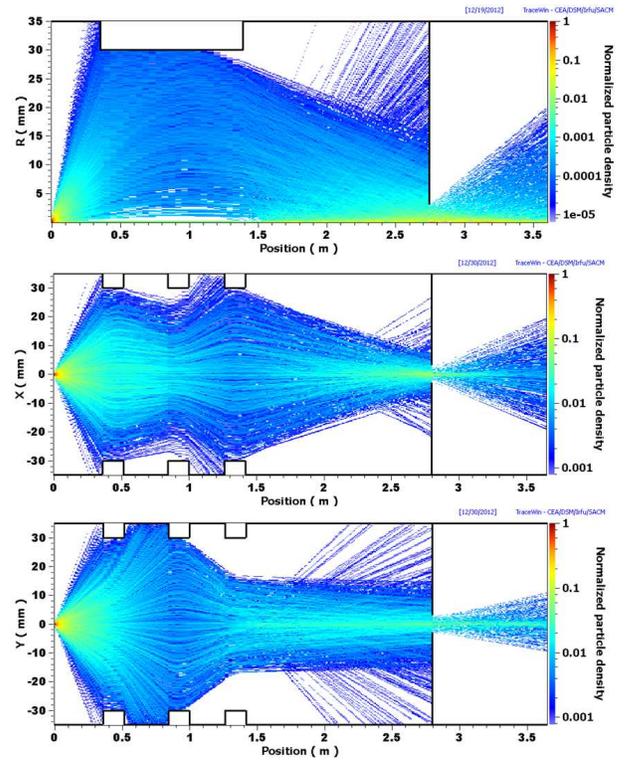}}
\caption{Density plots  for RPA-distribution in equivalent
solenoid (top) and triplet (center:x and bottom:y) systems
adjusted to 220 MeV.} \label{rpa-sol-triplet}
\end{figure}

\begin{figure}[h]
\centering\resizebox{0.45\textwidth}{!}{\includegraphics{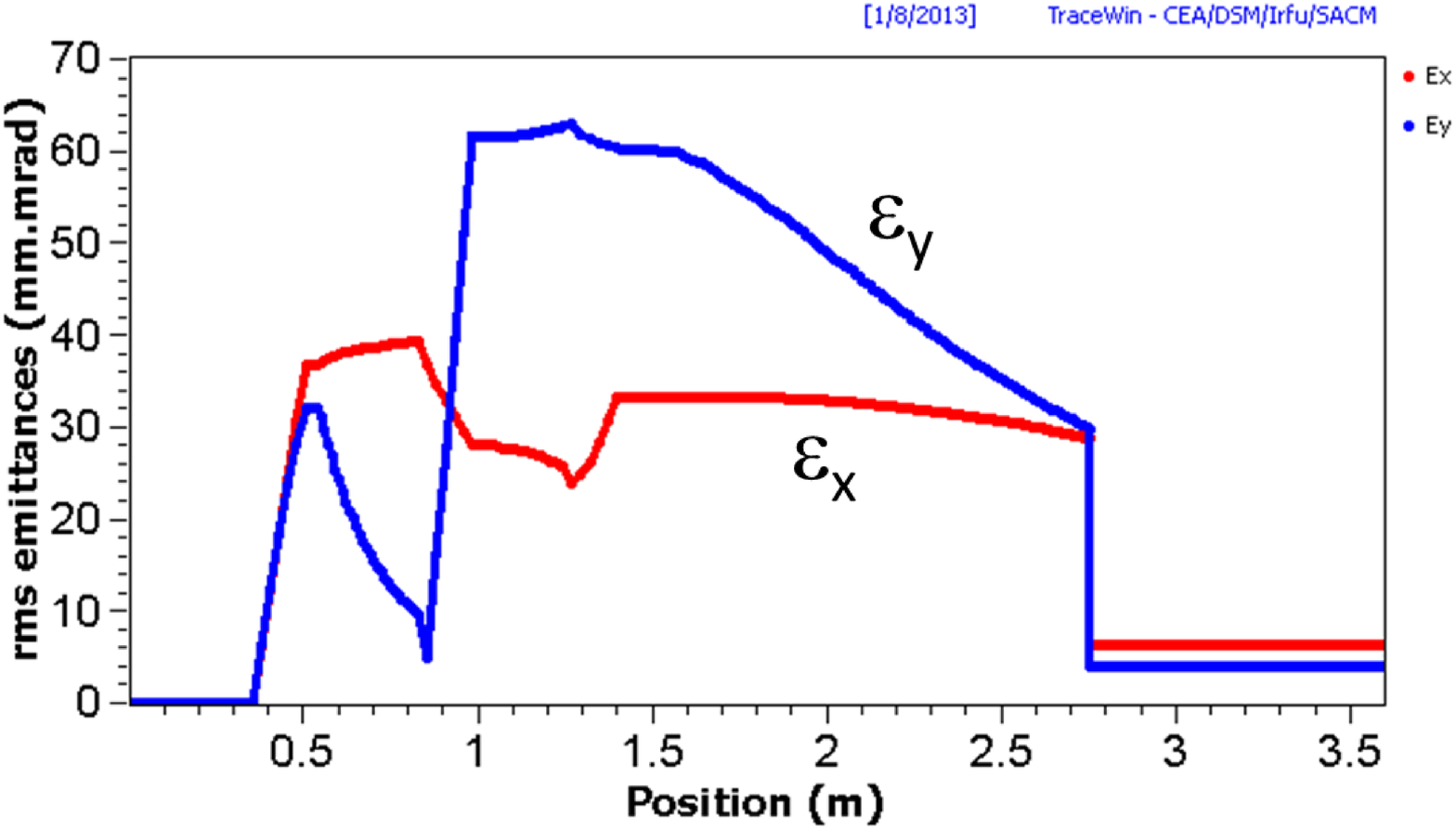}}
\caption{Transverse (normalized and bunch averaged) rms emittances
for triplet case.} \label{emit-triplet}
\end{figure}

\begin{figure}[h]
\centering\resizebox{0.45\textwidth}{!}{\includegraphics{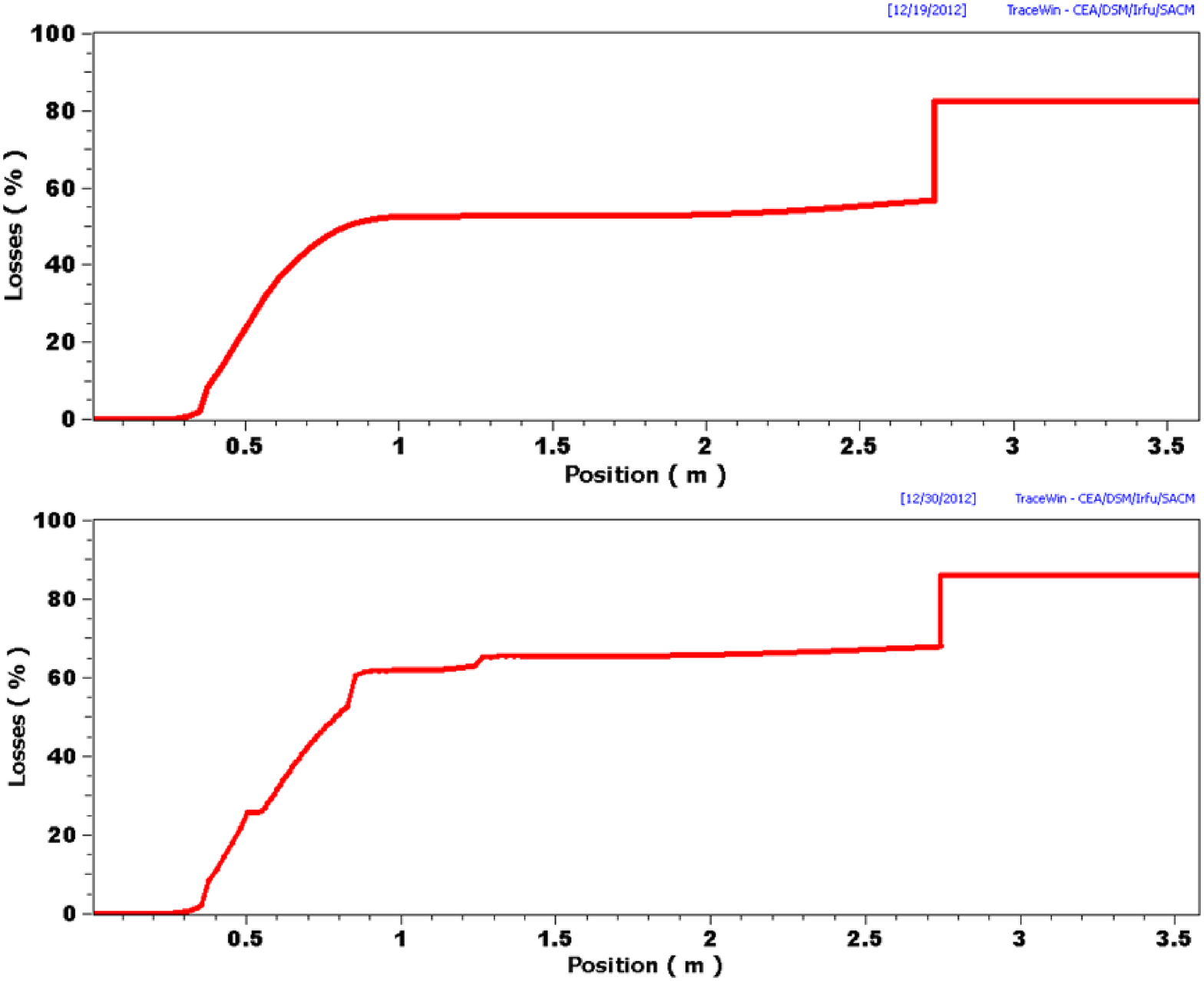}}
\caption{Loss profiles for RPA-distributions in the equivalent
solenoid (top) and triplet (bottom).} \label{loss-sol-triplet}
\end{figure}

\begin{figure}[h]
\centering\resizebox{0.45\textwidth}{!}{\includegraphics{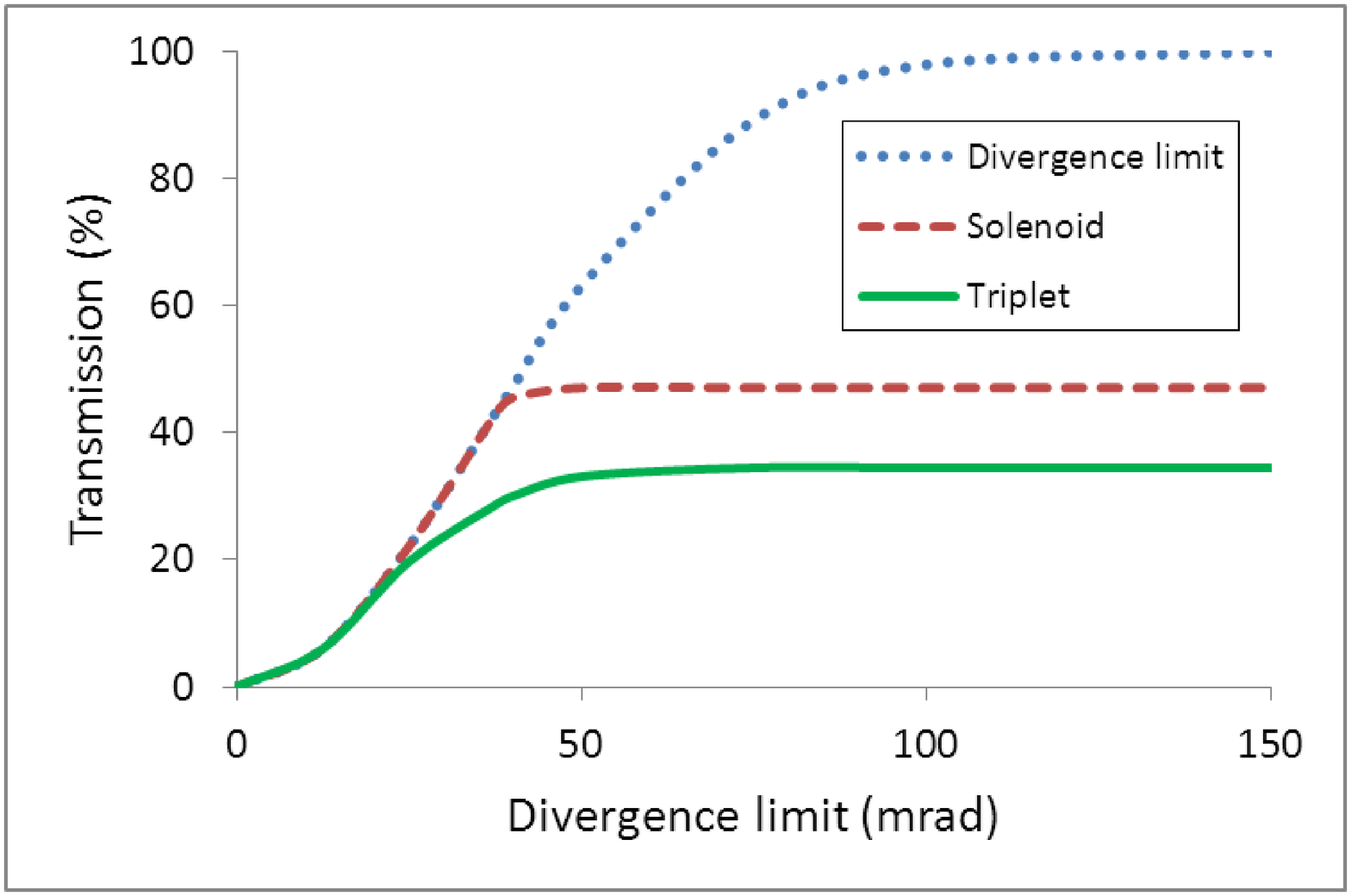}}
\caption{Transmission of solenoid and triplet as function of an
upper cut-off (divergence limit) of the injected particles. The
transmission is in \% of the particles in the original
un-truncated distribution.} \label{transmission}
\end{figure}

For this purpose we reduce the magnet fields in the equivalent
solenoid and triplet systems of Fig.~\ref{sol-triplet} for nominal
transmission at 220 MeV, which is closer to the peak of the energy
spectrum. We also require - arbitrarily - a focus at the distance
of 2.73 m from the laser target, where the energy selection
aperture is placed. Employing the above defined RPA-distribution
and a 3 mm radius aperture,   the resulting orbits of a TRACEWIN
simulation with 3000 rays  are shown in
Fig.~\ref{rpa-sol-triplet}.
In  Fig.~\ref{emit-triplet}  we examine the transverse emittances
for the triplet of Fig.~\ref{rpa-sol-triplet}. The relatively
large spread in $x', y'$ and energy width together with the energy
dependent focusing result in significant emittance growth  in $x$
and $y$ within the quadrupoles, accompanied by emittance
reductions due to beam loss on the radial aperture. Note that
emittances are understood here as averaged over the full bunch
length, while the ``instantaneous'' emittances at a given position
along the expanding bunch remain at their small initial values.
The emittance reduction in $y$ within the first quadrupole
reflects the beam loss in the defocusing $y-$direction.
A comparison of the equivalent solenoid and triplet focusing
systems shows that the overall transmission for the solenoid is
47\%, and 35\% for the triplet. The corresponding loss profiles
are shown in Fig.~\ref{loss-sol-triplet}.
The beam loss in $y$ in the first lens of the triplet is to some
extent compensated by an enhanced transmission in $x$, which
explains why the triplet transmission is only 25\% lower than the
solenoid one.   A more differentiated insight is gained, if we
truncate the divergence of the initial Gaussian distribution in
$x', y'$ by eliminating all particles beyond a ``divergence
limit'' as shown in Fig.~\ref{transmission}. Below about 40 mrad
the solenoid accepts all injected particles in the truncated
distribution (intensity in it relative to un-truncated
distribution indicated by the dotted line); at this value the
acceptance limit is reached and larger divergence particles are
lost. The triplet, instead, starts losing particles above 20 mrad,
but the transition to its acceptance limit is smoother -
apparently due to the benefit from the horizontal plane.

\subsection{Selection of energies}\label{eselect}
\begin{figure}[h]
\centering\resizebox{0.45\textwidth}{!}{\includegraphics{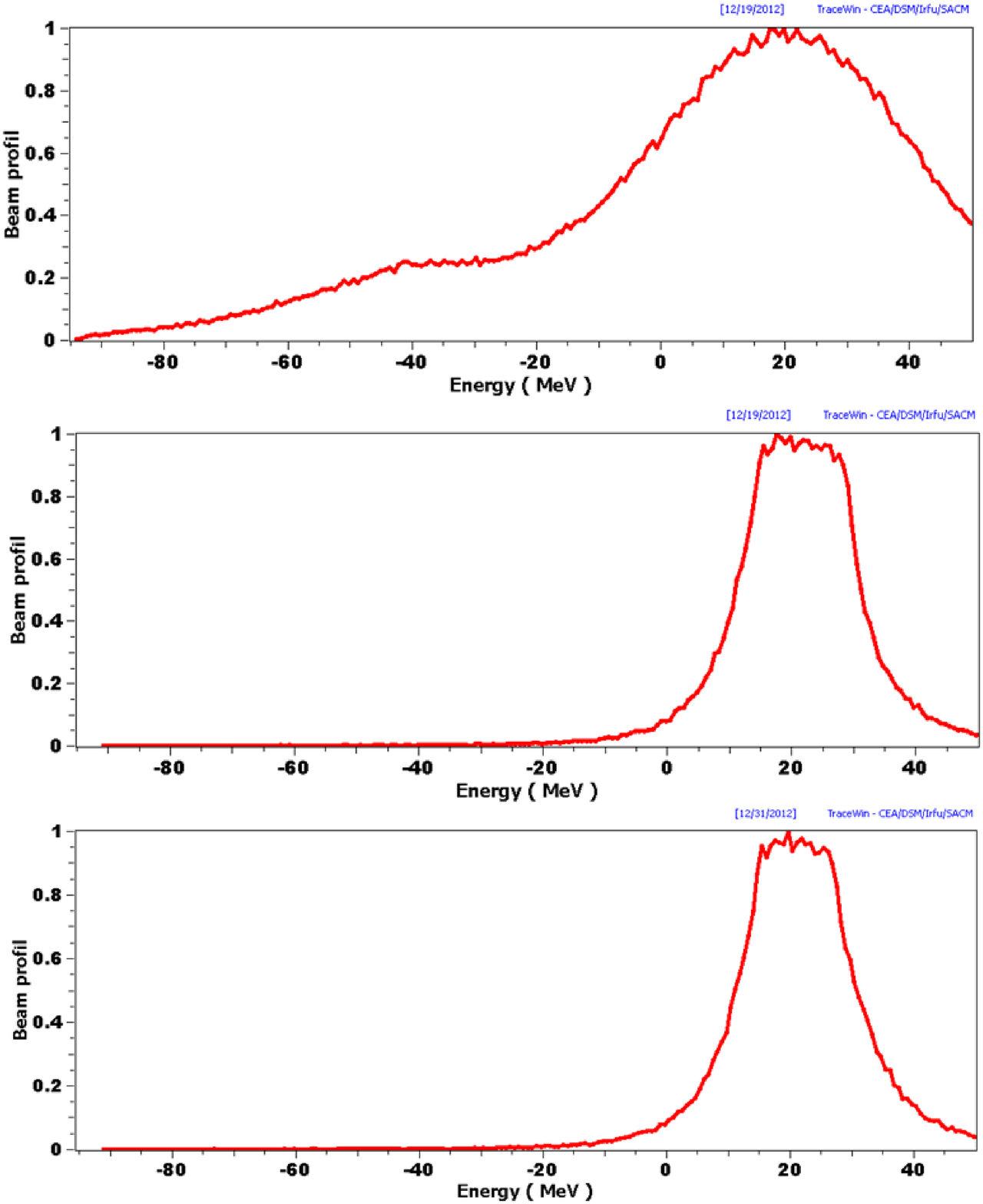}}
\caption{Selected energy spectra for equivalent solenoid
($R_{A}=3$ mm) and triplet systems ($R_{A}=2.7$ mm). Top: initial;
center: solenoid; bottom: triplet.} \label{energies}
\end{figure}

Following Ref.~\cite{hofmann2012a} the radius of a selection
aperture is proportional to the product of required energy width
and chromatic coefficient $\alpha$,
\begin{equation}\label{aperture}
    R_{A}=\alpha\frac{\Delta E}{E}A_{max},
\end{equation}
where $A_{max}$ is the maximum envelope at the lens. Using
$A_{max}\approx 3$ cm and $\alpha\approx 2$ we   expect that an
energy width of $\pm 4\%$ ($\pm 8.8$ MeV) should be obtainable
with an aperture of 2.4 mm radius. This is approximately confirmed
by the energy spectra in Fig.~\ref{energies}, where $R_{A}$ was
chosen as 3 mm for the solenoid and 2.7 mm for the triplet to
reach the same fwhm width.
The overall yield in the selected energy windows is 17\% for the
solenoid and 13 \% for the triplet, which follows approximately
the 25\% triplet transmission reduction found in
Fig.~\ref{transmission}.

For the triplet we have also simulated an elliptical selection
aperture in $x, y$ with the same area but semi-axes in the ratio
1:4 to match the ratio of chromatic coefficients according to
Eq.~\ref{aperture}. However, we obtain practically the same energy
selection width and profile as well as transmission. This seems
somewhat unexpected, but apparently the loss of selection in one
plane is compensated by better selection in the other plane.

\section{Conclusion}\label{sec:5}
The purpose of this study has been a comparative assessment of the
focusing properties of a solenoid and a quadrupole triplet in the
context of laser accelerated protons (or ions).  Possible
application of this acceleration method can be envisioned in the
field of particle therapy; but also in  proton radiography in
areas, where energies of a few hundred MeV, moderate integrated
intensities but high peak intensities in time-scales of few ns or
sub-ns are needed.  The relatively large initial angular and
energy spreads are a challenge  for all of these applications.

In terms of transmission it is found that ``equivalent'' systems -
same geometrical length and apertures - give only the relatively
small reduction in transmission  of 25\% for the triplet vs. the
solenoid, which is owed to the un-symmetric focusing of
quadrupoles. For increasing energies - already above a few MeV -
the weaker focusing properties of solenoids require pulsed or
super-conducting technology, whereas pole-tip fields of a doublet
or triplet can remain within room-temperature iron saturation.
This appears to be a clear advantage for quadrupoles in future
therapy applications, where short-term and fully controlled
changes of energy and magnetic rigidity are required.

The large energy spreads lead to a dominance of chromatic effects,
which can be used for energy selection.  Equivalent solenoid and
triplet systems are equally suitable for this purpose in spite of
the strongly differing chromatic coefficients in $x$ and $y$.
Chromatic effects lead to inevitable correlations between energy
and transverse position, which cannot be ignored for therapy
applications. In Ref.~\cite{hofmann2012a} it is shown that
properly placed scatter targets can be used to remove these
correlations.

The role of space charge and geometric aberrations - dominant for
short solenoids - needs further consideration even though they are
not expected to alter the major conclusions. Noting that solenoid
focusing is independent of the charge, the neutralizing co-moving
electrons, which are always present in laser acceleration, will be
strongly focussed towards the axis~\cite{almomani}. In the
triplet, instead, these electrons will be de-focussed to the
aperture when entering the first quadrupole, which may have an
effect on the quality of focusing and needs to be further
explored.

 \end{document}